# THE $^{59}$Fe(n, γ)$^{60}$Fe CROSS SECTION FROM THE SURROGATE RATIO METHOD AND ITS EFFECT ON THE $^{60}$Fe NUCLEOSYNTHESIS

S. Q. Yan(颜胜权)[1], X. Y. Li (李鑫悦)[1], K. Nishio[2], M. Lugaro[3,4,5], Z. H. Li(李志宏)[1,8], H. Makii[2], M. Pignatari[3,11,12,13], Y. B. Wang(王友宝)[1], R. Orlandi[2], K. Hirose[2], K. Tsukada[2], P. Mohr[6], G. S. Li (李广顺)[7], J. G. Wang (王建国)[7], B. S. Gao (高丙水)[7], Y. L. Han(韩银录)[1], B. Guo (郭冰)[1], Y. J. Li (李云居)[1], Y. P. Shen (谌阳平)[1], T. K. Sato[2], Y. Ito[2], F. Suzaki[2], J. Su(苏俊)[10], Y. Y. Yang (杨彦云)[7,8], J. S. Wang (王建松)[7], J. B. Ma (马军兵)[7], P. Ma (马朋)[7], Z. Bai (白真)[7], S. W. Xu (许世伟)[7], J. Ren (任杰)[1], Q. W. Fan(樊启文)[1], S. Zeng ( 曾晟)[1], Z. Y. Han (韩治宇)[1], W. Nan (南巍)[1], W. K. Nan (南威克)[1], C. Chen (陈晨)[1], G. Lian (连钢)[1], Q. Hu (胡强)[7], F. F. Duan (段芳芳)[7,9], S. Y. Jin (金树亚)[7,8,9], X. D. Tang (唐晓东)[7], and W. P. Liu(柳卫平)[1]

[1]China Institute of Atomic Energy, P. O. Box 275(10), Beijing 102413, P. R. China; panyu@ciae.ac.cn, wpliu@ciae.ac.cn
[2]Japan Atomic Energy Agency, Tokai, Ibaraki 319-1195, Japan
[3]Konkoly Observatory, Research Centre for Astronomy and Earth Sciences, Eötvös Loránd Research Network(ELKH), Konkoly Thege Miklós ùt 15-17, 1121 Budapest, Hungary; maria.lugaro@csfk.org
[4]ELTE Eötvös Loránd University, Institute of Physics, Budapest 1117, Pázmány Péter sétány 1/A, Hungary
[5]School of Physics and Astronomy, Monash University, VIC 3800, Australia
[6]Institute for Nuclear Research (ATOMKI), H-4001 Debrecen, Hungary
[7]Institute of Modern Physics, Chinese Academy of Sciences, Lanzhou 730000, China
[8]School of Nuclear Science and Technology, University of Chinese Academy of Sciences, Beijing, 100080, China
[9]School of Nuclear Science and Technology, Lanzhou University, Lanzhou 730000, China
[10]Beijing Normal University, Beijing 100875, P. R. China
[11]E.A. Milne Centre for Astrophysics, Dept of Mathematics and Physics, University of Hull, HU6 7RX, United Kingdom
[12]NuGrid Collaboration, (http://nugridstars.org)
[13]Joint Institute for Nuclear Astrophysics, Center for the Evolution of the Elements, Michigan State University 640 South Shaw Lane, East Lansing, MI 48824, USA

## ABSTRACT

The long-lived $^{60}$Fe (with a half-life of 2.62 Myr) is a crucial diagnostic of active nucleosynthesis in the Milky Way galaxy and in supernovae near the solar system. The neutron-capture reaction $^{59}$Fe(n,γ)$^{60}$Fe on $^{59}$Fe (half-life = 44.5 days) is the key reaction for the production of $^{60}$Fe in massive stars. This reaction cross section has been previously constrained by the Coulomb dissociation experiment, which offered partial constraint on the $E1$ γ-ray strength function but a negligible constraint on the $M1$ and $E2$ components. In this work, for the first time, we use the surrogate ratio method to experimentally determine the $^{59}$Fe(n,γ)$^{60}$Fe cross sections in which all the components are included. We derived a Maxwellian-averaged cross section of 27.5 ± 3.5 mb at $kT$= 30 keV and 13.4 ± 1.7 mb at $kT$= 90 keV, roughly 10 - 20% higher than previous estimates. We analyzed the impact of our new reaction rates in nucleosynthesis models of massive stars and found that uncertainties in the production of $^{60}$Fe from the $^{59}$Fe(n,γ)$^{60}$Fe rate are at most of 25%. We conclude that stellar physics uncertainties now play a major role in the accurate evaluation of the stellar production of $^{60}$Fe.

## 1. INTRODUCTION

The radioactive isotope $^{60}$Fe with a half-life of 2.62 Myr (Rugel et al. 2009; Wallner et al. 2015; Ostdiek et al. 2017) has been of interest to the nuclear physics and astrophysics communities for several decades. In our galaxy, the presence of $^{60}$Fe in the interstellar medium was confirmed through the detection of the 1173 and 1332 keV γ-rays from the decay of its daughter $^{60}$Co ($t_{1/2}$= 5.27 yr) by the RHESSI (Smith. 2003) and INTEGRAL satellites (Harris et al. 2005; Wang et al. 2007; Diehl 2013). Because the half-life of $^{60}$Fe is much shorter than the age of the galaxy, these observations provide the evidence of ongoing stellar nucleosynthesis. $^{60}$Fe has also been observed to be present in deep ocean ferromanganese crusts, nodules, sediments, snow from Antarctica (Knie et al. 1999, 2004;

Fitoussi et al. 2008; Wallner et al. 2016; Ludwig et al. 2016; Koll et al. 2019), and even in lunar regolith (Fimiani et al. 2016), which indicate one or more nearby supernova events occurred in the past several million years. Furthermore, $^{60}$Ni excesses are found in meteoritic materials, which indicate that $^{60}$Fe nuclei were present in the protoplanetary disk and may provide crucial information about the stellar environment of the nascent solar system (Shukolyukov & Lugmair 1993; Mostefaoui et al. 2004; Baker et al. 2005; Mishra & Goswami 2014; Telus et al. 2016, 2018; Trappitsch et al. 2018).

$^{60}$Fe is mainly produced in massive stars (M $\geqslant 8$ $M_\odot$) through neutron-capture reactions in the high neutron fluxes reached during C-shell burning and in the following explosive C burning and explosive He burning during the core-collapse supernova (CCSN) explosion (Limongi & Chieffi 2006; Jones et al. 2019). On the nucleosynthesis path of $^{60}$Fe production, the stable Fe isotopes capture neutrons until the unstable $^{59}$Fe is produced. Because the half-life of $^{59}$Fe is only 44.5 days, the production rate of $^{60}$Fe depends on the competition between neutron capture and the $\beta^-$ decay of $^{59}$Fe. The main neutron donor is the $^{22}$Ne($\alpha,n$)$^{25}$Mg reaction, and the neutron density is larger than $10^{11}$ neutrons cm$^{-3}$ (Limongi & Chieffi 2006). Accordingly, neutron capture dominates over $\beta^-$ decay, and $^{60}$Fe is produced in a substantial amount. At the same time, the produced $^{60}$Fe are destroyed by the $^{60}$Fe($n, \gamma$)$^{61}$Fe reaction.

To elucidate the production of $^{60}$Fe in massive stars, an accurate knowledge of the $^{59}$Fe(n, γ)$^{60}$Fe and $^{60}$Fe(n, γ)$^{61}$Fe reactions is necessary. While the Maxwellian-averaged cross section (MACS) of the $^{60}$Fe(n, γ)$^{61}$Fe reaction was experimentally determined to be 9.9 mb at $kT$ = 25 keV (Uberseder et al. 2009), no experimental data were available for the $^{59}$Fe(n, γ)$^{60}$Fe reaction until 2014 because of the difficulty in producing a short-lived $^{59}$Fe target for the direct measurement. In 2014, the Coulomb dissociation of $^{60}$Fe + Pb was used to constrain the $E$1 γ-ray strength function, and then the $^{59}$Fe(n, γ)$^{60}$Fe cross section was determined reversely (Uberseder et al. 2014). This experiment provided a pioneering constraint for the stellar nucleosynthesis of $^{60}$Fe. Nonetheless, because Coulomb dissociation populates the excited states of $^{60}$Fe by exciting the ground-state nuclei, the obtained $^{60}$Fe(γ$_o$,n)$^{59}$Fe cross sections offered partial constraint on $E$1 (Utsunomiya et al. 2010) and negligible constraint on the $M$1 or $E$2 γ-ray strength function of the $^{59}$Fe(n, γ)$^{60}$Fe, which caused a potential uncertainty in the determination of the cross section. Furthermore, the contribution of the $M$1 component was recently evaluated to be significant or even comparable to that of the $E$1 component (Loens et al. 2012; Mumpower et al. 2017). It follows that the rate is still very uncertain and recent studies have considered a potential variation of up to a factor of 10, with a strong effect on the model predictions (Jones et al. 2019). In this work, for the first time, we use the surrogate ratio method (SRM) (Escher et al. 2012) to experimentally determine the $^{59}$Fe(n,γ)$^{60}$Fe cross sections, which allow us to investigate all the components. Using this method, we measured the γ-decay probability ratios of the compound nuclei (CN) $^{60}$Fe* and $^{58}$Fe*, which were populated by the two-neutron transfer reactions of $^{58}$Fe($^{18}$O,$^{16}$O) and $^{56}$Fe($^{18}$O,$^{16}$O), respectively. Subsequently, the $^{59}$Fe(n, γ)$^{60}$Fe cross sections were determined using the measured ratios and the directly measured $^{57}$Fe(n, γ)$^{58}$Fe cross sections. We then tested the impact of our new rate in nucleosynthesis models of massive stars.

## 2. THE $^{59}$Fe(n,γ)$^{60}$Fe CROSS SECTION

### 2.1. *The Surrogate Ratio Method*

The SRM is a variation of the surrogate method (Younes & Britt 2003a, b; Petit et al. 2004; Boyer et al. 2006; Kessedjian et al. 2010). The method has been successfully employed to determine (n,f) cross sections (Plettner et al. 2005; Burke et al. 2006; Lyles et al. 2007; Nayak et al. 2008; Lesher et al. 2009; Goldblum et al. 2009; Ressler et al. 2011) and has recently been applied also to (n, γ) cross section measurements. A comprehensive review can be found in Escher et al. (2012), including both the absolute surrogate method and relative ratio method.

In this work, we determined the $^{59}$Fe(n, γ)$^{60}$Fe reaction cross section using the $^{57}$Fe(n, γ)$^{58}$Fe cross section according to the following equation:

$$\frac{\sigma_{^{59}Fe(n,\gamma)}(E_n)}{\sigma_{^{57}Fe(n,\gamma)}(E_n)} \approx C_{nor} \frac{N_{\gamma(^{60}Fe^*)}(E_n)}{N_{\gamma(^{58}Fe^*)}(E_n)}. \quad (1)$$

The derivation of this equation is described in Yan et al. (2016, 2017). The normalization factor $C_{nor}$ can be evaluated using the target thickness, the accumulated beam dose, and the γ-ray efficiency of the two surrogate reactions. $N_{\gamma(Fe^*)}(E_n)$ is the observed number of CN that decay into the ground state by emitting γ-rays, where $E_n$ is the equivalent neutron energy that yields the same excitation energy above the neutron sepa-ration energy. From the values of $N_{\gamma(^{60}Fe^*)}(E_n)$ and $N_{\gamma(^{58}Fe^*)}(E_n)$, the cross sections of the $^{59}$Fe(n,γ)$^{60}$Fe reaction can be determined using the known cross section of the $^{57}$Fe(n,γ)$^{58}$Fe reaction.

### 2.2. Benchmark Experiment

To check the validity of SRM for determining the (n, γ) cross section using the ($^{18}$O,$^{16}$O) surrogate reactions, we conducted a benchmark experiment to determine the $^{93}$Zr(n, γ)$^{94}$Zr cross sections (Yan et al. 2016) at astrophysical energies. The SRM-deduced cross sections agreed well with the directly measured cross sections. Furthermore, the neutron-capture cross section of the short-lived nucleus $^{95}$Zr (with half-life of 64 days) has been successfully determined to constrain the masses and metallicities of asymptotic giant branch stars where the meteoritic stardust SiC grains were born (Yan et al. 2017).

### 2.3. Measurement

The experiment was performed at the Tandem Accelerator of the Japan Atomic Energy Agency (JAEA) in Tokai. An $^{18}$O beam with an energy of 103.0 MeV was impinged onto an isotopically enriched iron target, which was prepared in the form of a self-supporting metallic foil. The thickness of the $^{56}$Fe target was 402 μg/cm$^2$ and the isotopic enrichment 99.4%. In the case of the $^{58}$Fe target, the thickness was 260 μg/cm$^2$ and the isotopic enrichment 96.3%. An array of ΔE – E silicon detector telescopes was located at the downstream of the target to identify the light ejectile particles, and four HPGe detectors were placed perpendicular to the beam direction at a distance of about 70 mm from the target for γ-ray detection. The absolute peak efficiency of each HPGe detector was about 0.6% at $E_γ$ = 1173.2 keV. A Faraday cup was installed about 1.3 meters away from the target to collect the $^{18}$O beam dose. The average intensity of $^{18}$O beam was about 0.2 pnA, and the diameter of the beam spot was less than 3 mm.

Each Fe target was irradiated for approximately 2.5 days, and the accumulated number of $^{16}$O was approximately $1.7 \times 10^5$ and $1.3 \times 10^5$ for the $^{56}$Fe and $^{58}$Fe tar- gets, respectively. The number of detected γ-ray events from $^{58}$Fe* and $^{60}$Fe* was about $4.5 \times 10^3$ and $3.4\times10^3$, respectively.

### 2.4. Data Analysis

The ejectile nucleus $^{16}$O from the $^{58}$Fe($^{18}$O, $^{16}$O)$^{60}$Fe* and $^{56}$Fe($^{18}$O, $^{16}$O)$^{58}$Fe* reactions was used to reconstruct the excitation energy $E_x$ of $^{60}$Fe* and $^{58}$Fe*, respectively, by two-body kinematics, and the γ-ray spectra from each CN were analyzed to obtain $N_{γ(^{60}Fe^*)}(E_n)$ and $N_{γ(^{58}Fe^*)}(E_n)$ in Eq. 1. To identify $^{16}$O, we used a two-dimensional scatter plot of energy loss (Δ$E$) versus total energy ($E_t$). Here, $E_t$ is the sum of energy loss in the Δ$E$ detector and the residual energy in the 16 strip annular $E$ detector. As an example, the Δ$E$ – $E_t$ scatter plot obtained from one of the combinations of Δ$E$ detectors and annular strips is shown in Fig. 1(a) with a cut to select $^{16}$O events from the ($^{18}$O,$^{16}$O) two-neutron transfer reaction. The energy resolution for $^{16}$O is about 0.5 MeV in FWHM, which is mainly due to the noise of the silicon detectors and the kinematic uncertainty due to the ∼0.6° acceptance of each ring.

The γ-ray spectrum obtained by gating the $^{16}$O region in the $^{18}$O + $^{58}$Fe reaction is shown in Fig. 1(b), where the 1290 keV $4^+ \to 2^+$ and 824 keV $2^+ \to 0^+$ transitions of $^{60}$Fe* are clearly observed. At the same time, $^{60}$Fe* exhibits a strong probability of neutron emission to yield $^{59}$Fe*, and the $^{59}$Fe* γ rays are evident in the spectrum. The 811 keV γ-ray corresponds to the $2^+ \to 0^+$ transition from $^{58}$Fe*, which indicates that a fraction of inelastic scattered $^{18}$O enter into the $^{16}$O gate.

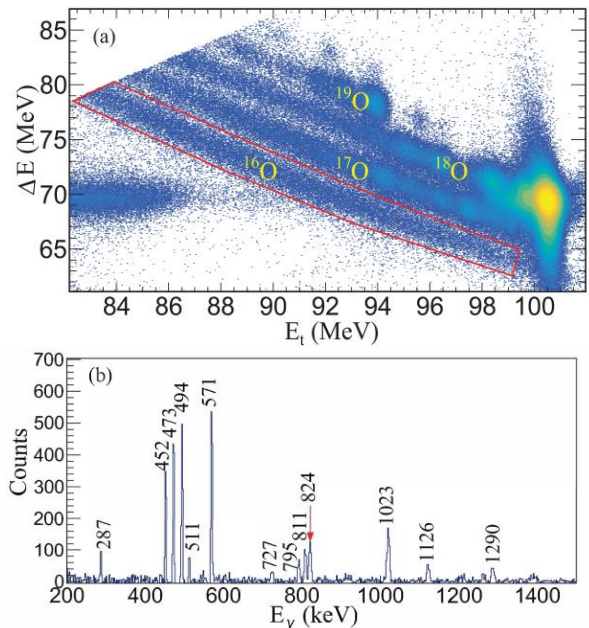

**Figure 1**. (a) Scatter plot of energy loss vs. total energy of the reaction products from $^{18}$O+$^{58}$Fe. The ($^{18}$O, $^{16}$O) events were selected using the gate shown in the plot. (b) γ-ray spectrum of $^{60}$Fe*, gated by $^{16}$O.

Because $^{60}$Fe and $^{58}$Fe are both even-even nuclei, the de-excitation of their high-lying resonance states is expected to overwhelmingly proceed through the doorway transition between the first excited $2^+$ state and the $0^+$ ground state. The energy of this transition is 824 keV for $^{60}$Fe*, and 811 keV for $^{58}$Fe*. In the analysis, the net areas were deduced from the 824 and 811 keV γ lines for each equivalent neutron-energy bin of $\Delta E_n$ = 500 keV. The net areas were then normalized to the integrated $^{18}$O beam dose, the target thickness, and the absolute detection efficiency of the HPGe detectors for each surrogate reaction. Based on the experimental data of HPGe γ detectors, the absolute branching ratios of 824 and 811 keV γ lines were obtained to be 86 ± 2% and 66 ± 3%, respectively. After the correction, we obtained the ratio of the γ-decay probabilities $N_{γ^{60}Fe^*}(E_n)/N_{γ^{58}Fe^*}(E_n)$. Considering the energy resolution in the equivalent neutron energy of 0.5 MeV, we obtained 16 values in the neutron-energy range of $E_n$ = 0-8 MeV.

### 2.5. Experimental Cross Sections

For applications to nuclear astrophysics, the (n, γ) cross sections are needed in the low-energy region of $E_n$< ~0.5 MeV. The desired low-energy cross section of the $^{59}$Fe(n,γ)$^{60}$Fe reaction can be calculated by the UNF (Zhang 1992, 1993, 2002) and TALYS (Koning et al. 2016) codes (using composite Gilbert-Cameron level densities and the Lorentzian model for the gamma-strength functions) after their level density parameter $a$ is constrained by the experimentally obtained γ-decay probability ratios in the high-energy region. According to Chiba and Iwamoto (Chiba & Iwamoto 2010), the γ-decay probabilities are relatively insensitive to the spin-parity distribution of CN at incident neutron energies $E_n$> ~ 3 MeV, and the γ-decay probabilities from different initial spin states tend to converge at high energies. Therefore, for the ratios of the γ-decay probabilities of two similar CN, e.g., like $^{60}$Fe* and $^{58}$Fe* in the present case, we can observe a good convergence with various spin parities in the high-energy region, which implies that the ratios obtained in surrogate experiments in the high-energy region are close to those obtained in neutron-capture measurements. Because the level density parameter $a$ is independent of the incident neutron energy, consequently, these high-energy experimental ratios can be used to constrain the parameter $a$, which in turn can be utilized to calculate the low-energy cross section.

The initial values of the theoretical input parameters for the UNF and TALYS codes were obtained from the RIPL (Capote et al. 2009) and TENDL (Koning et al. 2019) libraries. To determine the parameter $a$ of the UNF or TALYS code for the $^{59}$Fe(n, γ)$^{60}$Fe reaction, the $a$ for $^{57}$Fe(n, γ)$^{58}$Fe was initially fixed by the best fit to the directly measured data at low-energy energy region, then the high-energy cross sections could be obtained with the uncertainty less than 8%. Among the directly measured $^{57}$Fe(n, γ)$^{58}$Fe cross sections available in the literature (Macklin et al. 1964; Rohr & Müller 1969; Beer & Spencer 1975; Rohr et al. 1983; Wang et al. 2010; Giubrone et al. 2014; Giubrone 2014), the values reported by Macklin et al. (1964) are much higher than the others, and those derived by Beer & Spencer (1975), Wang et al. (2010) and Giubrone (2014) are consistent with each other. Considering the energy resolution, relatively accurate resonances, and higher-energy range, we used the latest data from Giubrone (2014). Because of the lack of the experimental data, the $^{60}$Fe giant dipole resonance parameters from systematics were used in UNF code: $σ_1$= 51 mb, $E_1$= 16.82 MeV, $Γ_1$= 4.33 MeV, $σ_2$= 45 mb, $E_2$= 20.09 MeV, $Γ_2$= 4.09 MeV. Then, the parameter $a$ was extracted ($a$= 7.807 MeV$^{-1}$, energy shift $\Delta$ = 0.05 MeV) from the best fit between the experimentally obtained ratios and the calculated cross section ratios at En= 3 - 8 MeV when the cross sections of the $^{59}$Fe(n, γ)$^{60}$Fe and $^{57}$Fe(n, γ)$^{58}$Fe reaction were calculated in high-energy region, as Fig. 2 shows. After the parameters were constrained, the low-energy cross sections of $^{59}$Fe(n, γ)$^{60}$Fe were calculated using the UNF and TALYS codes, the results are shown in Fig. 3. The uncertainty due to ratio fitting is about 8%, combining with the difference of 9% of the calculated cross section between the UNF and TALYS codes and uncertainty of the calculated $^{57}$Fe(n, γ)$^{58}$Fe cross section in high energy region, the cross section of $^{59}$Fe(n, γ)$^{60}$Fe reaction was determined with an uncertainty about 12% at $E_n$< 0.5 MeV.

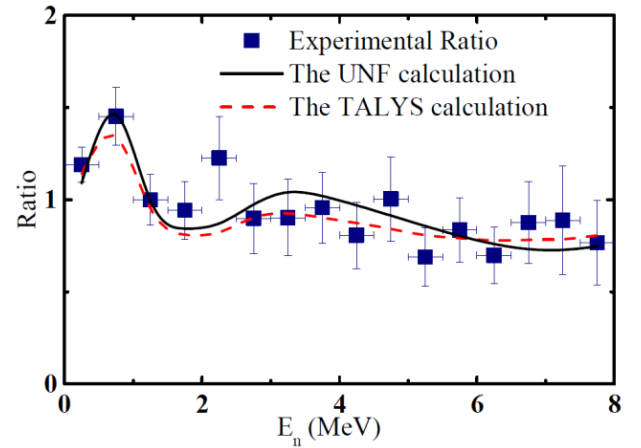

**Figure 2.** The γ-decay probability ratios of the compound nuclei $^{60}$Fe* and $^{58}$Fe*. The squares are the ratios obtained by the surrogate experiments. The dashed lines and the solid lines are the calculated (cross-section) ratios of the $^{59}$Fe(n,γ)$^{60}$Fe and $^{57}$Fe(n,γ)$^{58}$Fe reactions by the TALYS and UNF codes, respectively.

The average $^{60}$Fe*/$^{58}$Fe* γ-decay probability ratio was found to be 1.19 ± 0.10 at $E_n$≤ 0.5 MeV. The low-energy cross sections of the $^{59}$Fe(n, γ)$^{60}$Fe reaction can be deduced by multiplying the average experimental ratio with the directly measured $^{57}$Fe(n, γ)$^{58}$Fe cross section (Giubrone

2014), assuming that the ratio of the two (n, γ) cross sections can be approximated to a constant in this energy region. However, because of the low-lying levels of 14 and 136 keV in $^{57}$Fe and 287 keV in $^{59}$Fe, the $^{57}$Fe(n, γ)$^{58}$Fe cross sections are reduced at $E_n$ = 14 - 287 keV due to the additional inelastic neutron emission of $^{58}$Fe$^*$, and the ratios of the two (n, γ) cross sections fluctuate with $E_n$ in this low-energy region. Therefore, we used the Hauser – Feshbach theory to estimate the fluctuation in the ratio. The cross sections determined for the $^{59}$Fe(n, γ)$^{60}$Fe reaction are shown in Fig. 3. The uncertainty in the determined cross section includes an experimental uncertainty of about 8.5%, a systematic uncertainty of 5%, and the uncertainties involved in the direct measurement of $^{57}$Fe(n, γ)$^{58}$Fe cross sections. Here, the estimated experimental uncertainty includes a statistical uncertainty of 6%, a 3.5% uncertainty of the γ-branch ratio, and a 5% uncertainty arising from the $^{16}$O and γ-ray gates in their spectra. In the SRM, the CN formation cross section ratio of two-neutron-capture reactions and the ratio of the CN yield in two surrogate reactions are set to 1 to simplify the SRM formula in Eq. 1, which will bring a systematic uncertainty to the determined cross section; the details can be found in Yan et al. (2016). In the present work, the minor difference in the CN yield of the two surrogate reactions was corrected with experimental data, and the corresponding uncertainty was then reduced, but a statistical uncertainty (< 4%) was considered in this correction. Including the differences of CN formation cross sections between $^{57}$Fe + n and $^{59}$Fe + n (< 3%), a systematic uncertainty of 5% was counted in the cross sections determined.

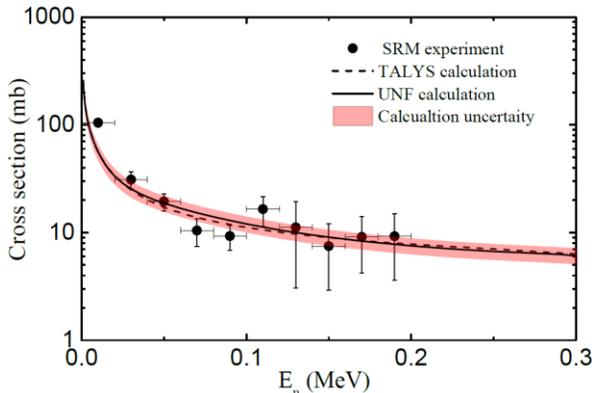

**Figure 3**. Variation in the cross section of $^{59}$Fe(n,γ)$^{60}$Fe as a function of equivalent neutron energy. The circles are obtained by multiplying the experimental γ-decay probability ratio with the directly measured $^{57}$Fe(n,γ)$^{58}$Fe cross section (Giubrone 2014). The dashed and solid curves represent the calculated results according to the UNF and TALYS codes, respectively, with their parameters constrained by the γ-decay probability ratios of CN $^{60}$Fe$^*$ and $^{58}$Fe$^*$ in the high-energy region.

## 2.6. Reaction Rate

After determining of the $^{59}$Fe(n, γ)$^{60}$Fe cross section, we derived the MACS at $kT$ = 40 - 100 keV for comparison with the results from the NON-SMOKER database (Rauscher & Thielemann 2000) and the Coulomb dissociation method, as Fig. 4 shows. The present MACS agree with the result of the Coulomb dissociation method within experimental uncertainties. The center value is almost 20% higher than that obtained by NON-SMOKER and almost 10% higher than that obtained in Uberseder et al. (2014). The present MACS for $^{59}$Fe(n,γ)$^{60}$Fe are shown in Table 1 for temperatures relevant to massive stars.

**Table 1**. Maxwellian-averaged cross sections of $^{59}$Fe(n,γ)$^{60}$Fe in mb.

| $kT$ [keV] | This work | Coulomb dissociation | NON-SMOKER |
|---|---|---|---|
| 30 | 27.5 ± 3.5 | - | 22.7 |
| 35 | 24.8 ± 3.2 | - | 20.5 |
| 40 | 22.6 ± 2.9 | - | 18.8 |
| 45 | 20.9 ± 2.7 | - | 17.3 |
| 50 | 19.5 ± 2.5 | - | 16.1 |
| 55 | 18.3 ± 2.3 | - | 15.0 |
| 60 | 17.4 ± 2.2 | - | 14.1 |
| 65 | 16.5 ± 2.1 | - | 13.3 |
| 70 | 15.7 ± 2.0 | - | 12.5 |
| 75 | 15.1 ± 1.9 | - | 11.9 |
| 80 | 14.5 ± 1.9 | $13.3^{+2.0}_{-3.1}$ | 11.2 |
| 85 | 13.9 ± 1.8 | $12.7^{+1.9}_{-3.0}$ | 10.8 |
| 90 | 13.5 ± 1.7 | $12.2^{+1.8}_{-2.9}$ | 10.3 |
| 85 | 13.0 ± 1.7 | $11.8^{+1.8}_{-2.8}$ | 10.0 |
| 80 | 12.6 ± 1.6 | $11.4^{+1.7}_{-2.8}$ | 9.6 |

In contrast to the Coulomb dissociation method where the $^{59}$Fe(n,γ)$^{60}$Fe cross section is obtained by constraining the upward γ-strength function of $E1$, we used SRM to measure the γ-decay probabilities of $^{60}$Fe$^*$ and $^{58}$Fe$^*$ and deduced the $^{59}$Fe(n,γ)$^{60}$Fe cross sections by including all the components. Because the contribution of the $M1$ component was shown separately in Uberseder et al. (2014), compared with the present MACS, we infer that the contribution of the $M1$ component to the total cross section is roughly 20%.

The corresponding reaction rate as a function of temperature $T_9$ (in units of $10^9$ K) is fitted with the expression used in the astrophysical reaction rate library RECLIB:

$$N_A \langle \sigma v \rangle = \exp(5.445 - 0.54 T_9^{-1} + 5.802 T_9^{-\frac{1}{3}} + 3.856 T_9^{-\frac{1}{3}} + 0.741 T_9 - 0.315 T_9^{\frac{5}{3}} - 0.18 \ln T_9)$$

The fitting errors are less than 5% in the range from $T_9$= 0.25 to $T_9$= 2.0.

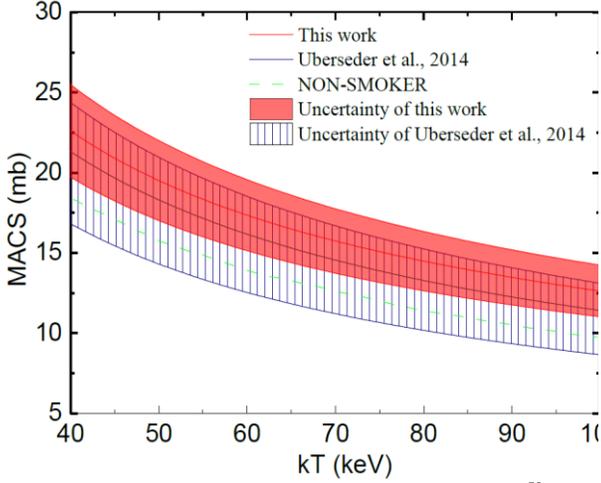

**Figure 4.** Comparison between the MACS of $^{59}$Fe(n, γ)$^{60}$Fe based on the present study, Ref. Uberseder et al. (2014), and the NON-SMOKER database (Rauscher & Thielemann 2000).

## 3  $^{60}$Fe PRODUCED IN MASSIVE STARS

We have tested the impact of the new $^{59}$Fe(n, γ)$^{60}$Fe rate presented here on the nucleosynthesis occurring in two models of massive star of initial masses 15 and 20 $M_\odot$ and solar metallicity (Z= 0.02) calculated by Ritter et al. (2018). For the latter phase of the 15 $M_\odot$ model, we tested two different setups for the convection-enhanced neutrino-driven explosion: fast-convection (the rapid setup) and delayed-convection (the delay setup) explosions (Fryer et al. 2012; Ritter et al. 2018). Because the results are very similar, we will mostly focus on the delay setup case, which we also used for the 20 $M_\odot$ model. To carry out the tests we used the NuGRID postprocessing code (Pignatari et al. 2016) and we calculated two sets of nucleosynthesis calculations: for one set we used the standard $^{59}$Fe(n, γ)$^{60}$Fe rate available in the JINA REACLIB database, version 1.1 (Cyburt et al. 2010), which is based on the NON-SMOKER HF model (Rauscher & Thielemann 2000). The second set is calculated by multiplying this rate by a constant factor of 1.66, consistent with the upper limit of the rate derived here. The rest of the nuclear reaction network is the same. Because the NON-SMOKER rate is similar to the lower limit derived here for the $^{59}$Fe(n,γ)$^{60}$Fe rate, with this test we can estimate the full impact of the new rate uncertainties on the stellar yield of $^{60}$Fe.

While both the pre-supernova hydrostatic and explosive components produce $^{60}$Fe via neutron captures through the $^{58}$Fe(n, γ)$^{59}$Fe(n, γ)$^{60}$Fe chain, the main region of production is different for the two components. In pre-supernova conditions, the bulk of $^{60}$Fe is made in the convective C shell. The CCSN explosion ejects a fraction of this $^{60}$Fe, while some part of it will be destroyed and produces new $^{60}$Fe by explosive C burning and explosive He burning. The relative importance of these different components in the total budget of the $^{60}$Fe yields may change between different stellar models depending on several parameters, the mass of the progenitor, and the explosion energy (e.g., Timmes et al. (1995); Limongi & Chieffi (2006); Jones et al. (2019)).

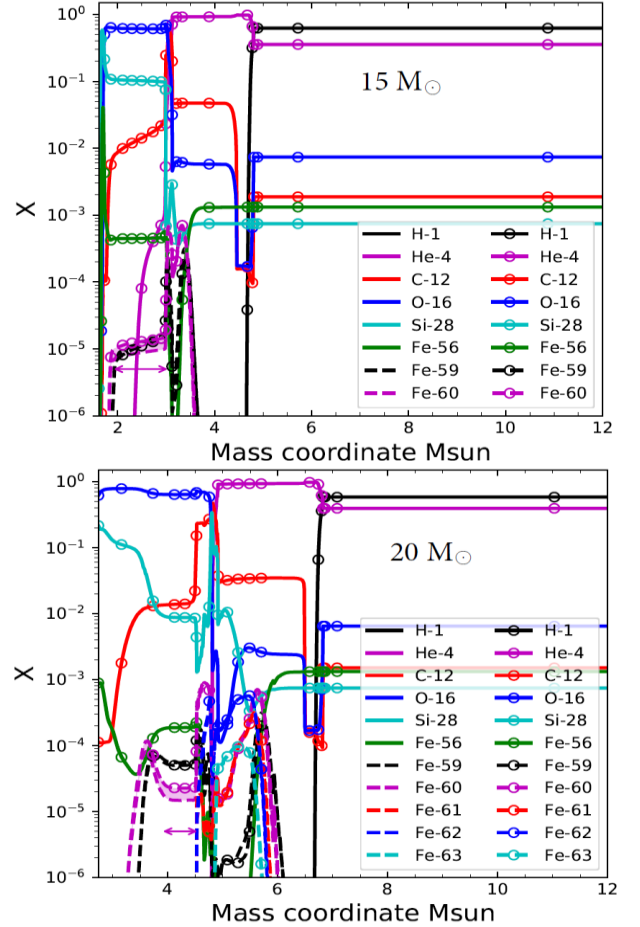

**Figure 5**. Abundance profiles in mass fraction (X) after the CCSN explosion of selected isotopes as function of internal mass, showing the impact of the $^{59}$Fe(n,γ)$^{60}$Fe rate on the production of $^{60}$Fe for the 15 $M_\odot$ (top panel) and 20 $M_\odot$ (bottom panel) models with the delay setup **(Ritter et al. 2018)**. For each isotope, lines with/without circles represent composition calculated using the standard NON-SMOKER×1.66/standard NON-SMOKER $^{59}$Fe(n,γ)$^{60}$Fe rate. The areas shaded in pink highlight the difference between the two $^{60}$Fe profiles. In the bottom panel also $^{61,62,63}$Fe are included, to highlight their production in the region of explosive He burning, together with $^{60}$Fe. Note that the progenitor of the 15 $M_\odot$ experienced a CO shell merger in the last days before collapse as noticeable by the high (0.1) $^{28}$Si abundance between mass 1.9 and 3 $M_\odot$.

In Figure 5, we show the abundance profiles for the ejecta of the 15 $M_\odot$ and the 20 $M_\odot$ models (top and bottom panels, respectively). The results obtained using the two different $^{59}$Fe(n, γ)$^{60}$Fe rates are compared. The top part of the C-shell burning ashes, at mass ranges of 3.9-4.6 $M_\odot$ (bottom panel) and 1.9-3.0 $M_\odot$ (top panel) about 1.9 $M_\odot$ (top panel) and 3.3 $M_\odot$ (bottom panel). This is a common feature of stars of mass in the range considered here. Depending on the explosion energy and on the progenitor structure, explosive C-burning can efficiently produced new $^{60}$Fe. For instance, in the bottom panel of Figure 5 at a mass coordinate of about 3.5 $M_\odot$ we obtain a peak of Fe, with some impact of the standard $^{59}$Fe(n, γ)$^{60}$Fe rate uncertainty. The 15 $M_\odot$ model instead does not show the signature of explosive C burning.

For both of the two models shown in Figure 5, the main 60Fe production is due to explosive He burning, as the peak just above mass 3 $M_\odot$ in the 15 $M_\odot$ model, and as the two peaks between 5 and 6 $M_\odot$ in the 20 $M_\odot$ model. For the explosive production of $^{60}$Fe in He-buring conditions, the difference between the two cases calculated with different $^{59}$Fe(n, γ)$^{60}$Fe rates is not significant. In fact, in Figure 5 there is no highlighted pink area here as for the pre-supernova C-burning ejecta. The reason for this becomes clear if we look at the abundance profiles shown for the 20 $M_\odot$ model, where more isotopes are reported along the neutron-capture chain from $^{59}$Fe to $^{63}$Fe. In explosive He-burning conditions, the neutron density rises quickly to values above $10^{18}$ neutrons per cm$^3$, typical of the neutron burst in explosive He-burning conditions (n-process, Meyer et al. 2000; Pignatari et al. 2018). In these conditions, neutron capture on $^{60}$Fe feeds the production of $^{61}$Fe and $^{62}$Fe. A smaller (higher) $^{59}$Fe(n, γ)$^{60}$Fe rate will reduce (increase) the production of $^{60}$Fe. On the other hand, a smaller (higher) quantity of $^{60}$Fe will be depleted less (more) efficiently to make more neutron-rich Fe isotopes. The balance between production and destruction of $^{60}$Fe causes its abundance to reach equilibrium and become less affected by the $^{59}$Fe(n, γ)$^{60}$Fe rate.

In Figure 6 we summarize the results for the five different nucleosynthetic environments: the pre-supernova models for both 15 $M_\odot$ and the 20 $M_\odot$ stars, the two CCSN explosive setups (the rapid setup and the delay setup) for the 15 $M_\odot$ model, and the one CCSN explosive setup for the 20 $M_\odot$ model. Overall, the impact of increasing the $^{59}$Fe(n, γ)$^{60}$Fe rate is much more significant during the hydrostatic phase, where variations in the final $^{60}$Fe yield are above 25% in the case of the 20 $M_\odot$ stars. After the explosion, in the models presented here the variation factor of total $^{60}$Fe yields decreases to less than 10%. This is due to the dominant contribution to the $^{60}$Fe made by explosive nucleosynthesis, compared to the ejecta with $^{60}$Fe made before the SN explosion (see Figure 5). Notice that in models with a weaker explosion than those presented here, the pre-supernova components would become more relevant, and the impact of variations in the $^{59}$Fe(n, γ)$^{60}$Fe on the final $^{60}$Fe yields would be quantitatively much closer to the values seen in the progenitor pre-supernova models. These results are in qualitative agreement with those presented by Jones et al. (2019). In that paper, when the $^{59}$Fe(n, )$^{60}$Fe rate was increased by a factor of 10, the preexplosive $^{60}$Fe yield increased but there was no further increase during the explosion.

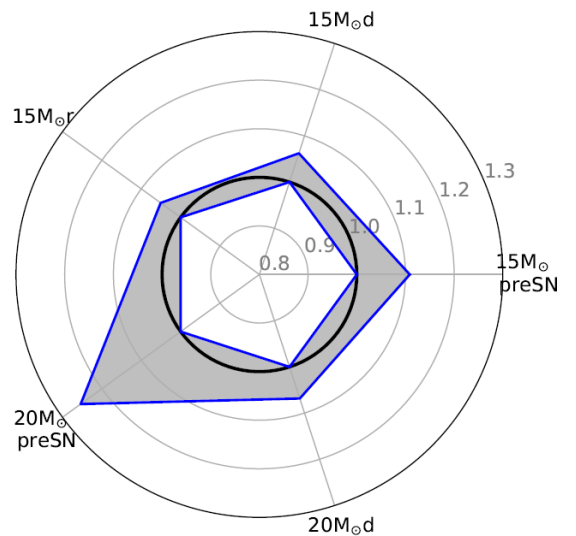

**Figure 6**. Summary of the effect of varying the $^{59}$Fe(n, γ)$^{60}$Fe rate, where variations in the $^{60}$Fe yields are indicated in the form of concentric circles, labelled with numbers representing the fraction between the yields calculated with the higher rate (i.e., the rate multiplied by 1.66, external blue shape) and the NON-SMOKER case (i.e., the NON-SMOKER rate, blue pentagon, touching the thick black line circle labelled as 1.0).

In summary, the production of $^{60}$Fe strongly depends on the progenitor evolution and on the explosion uncertainties. In particular, if we exclude a 15 $M_\odot$ outlying CCSN model at high explosion energy, the Jones et al. (2019) yields showed a variation in $^{60}$Fe production more than an order of magnitude. This result was obtained considering a range of explosion energies between a few $10^{50}$ erg and $5 \times 10^{51}$ erg, and three stellar progenitor masses. Such a variation contributed by stellar physics such as the explosion energies is a factor of 3 if we compare CCSN models from the same stellar progenitors and with SN explosion energies smaller than $2 \times 10^{51}$ erg (Jones et al. 2019). In our work we show that the impact of the $^{59}$Fe(n, γ)$^{60}$Fe uncertainty in preexplosive yields provides an upper limit of the variation in the final postexplosive yields. With the present $^{59}$Fe(n, γ)$^{60}$Fe errors, in our models, the largest impact on the $^{60}$Fe yields that we obtain for this reaction rate is within a 30% variation.

## 4. SUMMARY AND CONCLUSION

In this work, we have overcome the outstanding experimental challenges in the measurement of $^{59}$Fe(n, γ)$^{60}$Fe cross sections. To the best of our knowledge, this is the first study that presents experimental constraints for all the components in these cross sections using SRM. We clarified the considerable uncertainties of the $M$1 and $E$2 components from Uberseder et al. (2014) and provided a complete MACS for studies of stellar production of $^{60}$Fe with the impact on ongoing galactic nucleosynthesis, nearby supernova events, and the history of our solar system. Based on the new rate presented here and the result of our modeling tests, the main uncertainties in the derivation of the $^{60}$Fe yields from massive stars are related to the stellar physics of the progenitor and of the subsequent supernova explosion, rather than to the value of the $^{59}$Fe(n, γ)$^{60}$Fe rate.

We thank the staff of the JAEA Tandem Accelerator facility for their help with the experiment. We also thank the staff of Radioactive Ion Beam Line in Lanzhou (RIBLL) for the feasibility test of this experiment. This work was supported by the National Key Research and Development Program of China under grant No. 2016YFA0400502, the National Natural Science Foundation of China under grant Nos. 11875324, 11961141003, and 11490560, the Continuous Basic Scientific Research Project No. WDJC-2019-13, the Leading Innovation Project of the CNNC under grant Nos. LC19220900071 and LC202309000201, NKFIH (Nos. K120666 and NN128072), New National Excellence Program of the Ministry for Innovation and Technology (No. ÚNKP-19-4-DE-65) and the ERC Consolidator Grant (Hungary) funding scheme (Project RADIOSTAR, G.A. n. 724560). We acknowledge significant support from NuGrid through NSF grant PHY-1430152 (JINA Center for the Evolution of the Elements) and STFC (through the University of Hull's Consolidated Grant ST/R000840/1), and access to viper, the University of Hull High Performance Computing Facility. We acknowledge the support from the "Lendület-2014" Programme of the Hungarian Academy of Sciences (Hungary). We also thank the UK network BRIDGCE and the ChETEC COST Action (CA16117), supported by the European Cooperation in Science and Technology, the ChETEC-INFRA project funded from the European Union's Horizon 2020 research and innovation programme under grant agreement No. 101008324, and the IReNA network supported by NSF AccelNet).